\documentclass[12pt]{article}
\usepackage{graphicx,amsmath,amssymb,color,bm}
\renewcommand{\baselinestretch}{2}

\begin{document}

\title {Electronic and optical properties of stacking-configuration-modulated bilayer graphene in electric and magnetic fields}
\author{
\small Chiun-Yan Lin$^{a}$, Ming-Fa Lin$^{a,b*}$ $$\\
\small  $^a$Department of Physics, National Cheng Kung University, Tainan 701, Taiwan \\
\small  $^b$Hierarchical Green-Energy Materials/quantum topology centers, \\
\small  National Cheng Kung University, Tainan 701, Taiwan }
\renewcommand{\baselinestretch}{1.66}
\maketitle

\renewcommand{\baselinestretch}{1.66}

\begin{abstract}

The electronic properties and optical excitations are investigated in the geometry- and field-modulated bilayer graphene systems, respectively, by using the tight-binding model and Kubo formula.
The stacking symmetry of bilayer graphene can be manipulated by varying the width and position of domain wall (DW) within two normally stacked graphene.
All the layer-dependent atomic interactions are taken into consideration under external fields.
The modulation of stacking configuration gives rise to significant effects of zone folding on energy subbands, subenvelope wave functions, density of states, and optical absorption spectra.
This study clearly illustrates the diverse 1D phenomena in the energy band structure and absorption spectra; the DW- and $V_z$-created dramatic variations are comprehensively explored under accurate calculations and delicate analysis.
Concise physical pictures are proposed to give further insight into the quasi-1D behaviors.

\end{abstract}

\par\noindent  * Corresponding author.
{~ Tel:~ +886-6-275-7575.}\\~{{\it E-mail addresses}:  mflin@mail.ncku.edu.tw (M.F. Lin)}

\pagebreak
\renewcommand{\baselinestretch}{2}
\newpage

\vskip 0.6 truecm

\section{Introduction}

Stacking configurations and external electric $\&$ magnetic fields can greatly diversify the essential physical properties. Three typical  categories of bilayer graphenes, the geometry-modulated, sliding and twisted systems, clearly display the artificial manipulations of stacking symmetries. The up-to-date experimental synthesis methods on the stacking modulations of bilayer graphene systems cover the mechanical exfoliation\cite{5Nature520;650}, and chemical vapor deposition\cite{5Nature505;533}, and tips of STM\cite{5NatComm7;11760}. Very important, the former is capable of tuning the width and position of domain wall (DW) between the normal stackings by manipulating atomic force microscopy (AFM) tip\cite{5NatNanotech13;204}. On the theoretical side, the first-principles method\cite{5PRX3;021018} and the tight-binding model\cite{5SciRep9;859} have been used to explore energy bands of the AB/DW/BA/DW bilayer graphene, with a very narrow domain-wall width [only comparable to one C-C bond length] and a middle period. Whether the stacking-modulated band structures present quasi-2D or quasi-1D will be clarified in this study. It is well known that a uniform perpendicular electric field leads to an opening of band gap in bilayer AB stacking, confirmed by the theoretical\cite{5IOP;CY} and experimental results\cite{5Nature459;820}. Periodical gate voltages [$V_z$'s], with the opposite Coulomb potentials in the neighboring ones, could be achieved under the delicately experimental designs\cite{5Nature459;820}. This will be very efficient in modulating the low-energy electronic properties. Furthermore, the theoretical calculations predict the existence of 1D topological states\cite{5PRX3;021018}.\\

As to bilayer graphene systems, there exist three types of well-behaved stacking configurations in bilayer graphenes\cite{5IOP;CY}. According to the low-lying electronic structures, AA, AA$^\prime$, and AB stackings, respectively, exhibit the vertical $\&$ non-vertical two Dirac-cone structures, and two pairs of parabolic bands. Such band structures are clearly illustrated along the high-symmetry points of the hexagonal first Brillouin zone, so that 2D behaviors are responsible for the other fundamental physical properties. For example, the density of states and optical spectra obviously reveal the 2D van Hove singularities, where the special structures are very sensitive to the dimension. Similar 2D phenomena appear in the sliding and twisted systems, while the important differences appear between them\cite{5SciRep9;859,5SciRep4;7509}. Random stacking configurations in the former show the highly distorted energy dispersions with an eye-shaped stateless region accompanied by saddle points. Furthermore, the Mori\'{e} superlattice of the latter possesses a lot of carbon atoms in a primitive the unit cell and thus creates many 2D energy subbands. Most importantly, these two systems are expected to be in sharp contrast with the geometry-modulated bilayer graphenes in any physical.\\

In this work, we first investigate electronic properties and optical excitations in the geometry- and electric-field-modulated bilayer graphene systems, respectively, by using the tight-binding model and Kubo formula in gradient approximation. The modulation of stacking configuration will lead to a great enlargement of the unit cell. The significant effects of zone folding on energy subbands, subenvelope functions on distinct sublattices, the density of states, and absorption structures are fully explored under the accurate calculations and delicate analysis. The concise physical pictures are proposed to account for the complex relations among the geometric modulation, the gate voltage and the 1D behaviors. Very important, the current study clearly illustrates the diverse 1D phenomena, a lot of energy subbands with various band-edge states, the metallic behavior even in an electric field, the unusual van Hove singularities, the forbidden vertical excitation channels under the specific linear relations of the layer-dependent sublattices, the prominent asymmetric absorption peaks in absence of selection rule, and the DW- and $V_z$-created dramatic variations in optical absorption structures. Obviously, the non-uniform magnetization due to the stacking modulation can only be solved by the generalized tight-binding model, but not the effective-mass approximation. Most importantly, the strong competition between the geometric symmetry and magnetic field is studied in detail. This method is very reliable even or the complicated energy bands with the oscillatory dispersions and the multi-constant loops.  The non-homogeneous interlayer hopping integrals and Peierls phases are simultaneously included in the diagonalization of the giant magnetic Hamiltonian matrix, where they have induced the high barriers in the numerical calculations. The magneto-electronic properties, energy spectrum, density of states, and wave functions, directly link one another by the competitive/cooperative pictures. The oscillatory Landau subbands are predicted to come to exist under the non-uniform environment, being thoroughly different from the dispersionless Landau levels in the well-stacked graphene systems\cite{5IOP;CY}. The above-mentioned significant 1D features are expected to sharply contrast with the 2D ones in the twisted \cite{5NanoLett12;3833} and sliding\cite{5SciRep4;7509} bilayer graphenes. Furthermore, they are thoroughly compared with the dimensionality-induced phenomena, e.g., the important differences of Landau subbands among the stacking-modulated bilayer graphenes\cite{5SciRep9;859}, graphene nanoribbons\cite{5PCCP18;7573}, and AB- and ABC-stacked graphites\cite{5CRCPress;CY}. Such theoretical predictions require a series of  experimental examinations from the STS, optical and magneto-optical spectroscopies.\\

\subsection{Electronic properties and absorption spectra}\label{ch5.1}

Electronic and optical properties of the AB/DW/BA/DW bilayer graphene are greatly diversified by the manipulation of geometry and gate voltage. The periodical modulation along $\hat{x}$ induces the dramatic 2D to 1D changes, covering a lot of energy subbands and well-behaved/irregular standing waves, various van Hove singularities in density of states (the double- and single-peak structures, a pair of rather strong peaks, and a plateau across $E_F$), and the metallic behavior. The optical gaps vanish in any systems. A pristine bilayer AB stacking exhibits the featureless optical spectrum at lower frequency. As to the geometry-modulated systems, the observable absorption peaks could survive only under the destruction of the symmetric/anti-symmetric linear superposition due to the interlayer (A1,A2)/(B1,B2) sublattices. The frequency, number and intensity of absorption structures are very sensitive to the modulation domain wall and Coulomb potential. \\

A bilayer graphene, as clearly shown in Fig. 1(a), is periodically modulated along the $\hat{x}$ direction, in which the translation symmetry in the $y$-axis remains unchanged. This special geometric structure covers the wide AB and BA stacking regions, respectively, at the left- and right-hand sides, and the domain walls (DWs) in between them. The slowly stacking transformation is assumed to appear in DWs along the armchair direction by a uniform variation of the C-C bond lengths. Obviously, random stacking configurations appear within the ranges of DWs, leading to the non-uniform environment and thus the destruction in most rotation symmetries. A typical period within (60,10,60,10) [in the unit of $3b$; with $b$ denoting the C-C bond length], with 1120 carbon atoms, is very suitable for a full exploration of the modulation-diversified essential properties. That is to say, the ${1120\times\,1120}$ Hamiltonian matrix, which covers the various intralayer and interlayer hopping integrals in Eq. (4.2) \cite{5SciRep9;859}, is available in fully exploring the gate-voltage-dependent electronic properties and optical absorption spectra. Such a bilayer system possesses a periodical boundary condition with a long period, so that the dimension-enriched electronic and optical properties will frequently come to exist. The total carbon atoms in a primitive unit cell can be classified into the four [A$^1$, B$^1$, A$^2$, B$^2$] sublattices. In the numerical calculations, the wave-vector-dependent wave functions strongly depend on the odd or even indices of each sublattice. The fundamental physical properties are very sensitive to the application of an external electric field/gate voltage [Fig. 1(b)]. $V_z$ can generate the layer-dependent Coulomb potentials and thus destroy the mirror symmetry about the $z=0$ plane.\\

The low-lying electronic structures of symmetric and asymmetric bilayer graphenes exhibit the unusual features. Obviously, a pristine bilayer AB stacking possesses two pairs of valence and conduction bands, with parabolic energy dispersions initiated from the K $\&$ K' valleys [details in Refs\cite{5IOP;CY,5CRCPress;CY}]. The first pair presents a slight overlap around $(2\pi/\sqrt{3}a,2\pi/3a)$ and $(2\pi/\sqrt{3}a,-2\pi/3a)$, where $a=\sqrt{3}b$ is the lattice constant of monolayer graphenes. Furthermore, the second pair appears at $\sim \pm 0.42$ eV, mainly owing to the dominance of the interlayer vertical hopping integral. The lower-energy bands of the former are chosen as the studying focus. In order to compare the main features of electronic structures between the pure and geometry-modulated bilayer systems, energy bands of the former, corresponding to an enlarged unit cell, are folded into a specific valley around $k_y a=2\pi/3$ [the K point in Fig. 2(a)]. That is, there are a lot of 1D parabolic dispersions due to the zone-folding effects. Since this system presents a translational invariance along the $\hat x$ direction, all valence and conduction bands are doubly degenerate except for those two bands nearest the Fermi level [Fig. 2(a)]. The particle-hole symmetry is obviously broken under the interlayer atomic interactions. According to state energies measured from $E_F$, the first, second and third pairs are denoted as $(v_1, c_1)$, $(v_2,c_2)$/$(v'_2, c'_2)$; $(v_3,c_3)$/$(v'_3 , c'_3)$, and so on. All band-edge states in distinct energy bands are very close to the $K$ point.\\

Specifically, the wave functions at the $K$ point, as indicated in Fig. 3(a), clearly illustrate the unique characteristics of the spatial distributions for various energy subbands. The subenvelope functions on four sublattices [$(A^1, B^1, A^2, B^2)$ by the red, green, black and blue colors, respectively] are responsible for four components of Bloch wave functions. The $v_1$ and $c_1$ states in Figs. 3(a) and 3(b), which have constant values for each component, are the anti-symmetric and symmetric superposition of only the A$^1$- and A$^2$-dependent tight-binding functions, respectively. Although $v_1$ and $c_1$ present the different dominating components, these two states are degenerate at the $K$ point [Fig. 2(a)] and might have other ratios between the weights of A$^1$- and A$^2$-sublattices through the linear combination. Notice that, due to armchair shape in the $x$-direction, each sublattice is further split into the odd and even indices. The eigenfunctions between two distinct indices within the same sublattice only exhibits a $\pi$ phase difference. The odd-index components will account for a clear presentation. With the increasing 1D subband indices, the wavefunctions become the well-behaved standing waves instead of the uniform spatial distributions. For $(v_2, c_2)$/$(v'_2, c'_2)$ in Figs. 3(c)-3(d)/Figs. 3(e)-3(f), the K states have the dominant components in the A$^1$- and A$^2$-sublattices and minor weights in B$^1$- and B$^2$-sublattices. Apparently, the four subenvelope functions show the standing-wave behaviors with two nodes in each component of the spatial distributions, arising from the linear combination of the $\pm k_x$ states, as mentioned earlier. The $\pi/2$-phase difference for each sublattice between $v_2$ and $v'_2$ [$c_2$ and $c'_2$] is consistent with a uniform distribution in a 2D pristine system. The calculated results display that the subenvelope functions in the A$^1$- and A$^2$-sublattices are almost in-phase for conduction bands [Figs. 3(d) and 3(f)] and out-of-phase for valence ones [Figs. 3(c) and 3(e)], and the opposite is true for those between the B$^1$- and B$^2$-sublattices. The slight phase shift between A$^1$ and A$^2$ or between B$^1$- and B$^2$ is intrinsic, which depends on calculation parameters. The phase shift becomes less obvious for modes with higher energy. As to the next energy subbands of $v_3$ and $c_3$, the eigenfunctions reveal the four-node standing waves in the well-behaved forms [Figs. 3(g) and 3(h)]. The number of nodes, which indicates the quantization behavior along $\hat{x}$, is expected to continuously grow in the increment of subband index. \\

Apparently, the drastic changes of band structures come to exist for the geometry-asymmetry bilayer graphene, as clearly indicated in Figs. 2(b) and 2(c). They cover band asymmetry, band overlap, state splitting, energy dispersions, band-edge states, significant subband hybridizations, and distorted standing-wave behaviors. The asymmetry of energy spectrum about $E_F=0$ becomes more obvious in the presence of the geometric modulation. This result is purely due to more non-uniform interlayer hopping integrals [details in Eq. (5.2)]. The overlap of valence and conduction energy bands is getting larger, and so do the free electron and hole densities. The destruction of the $(x,y)$-plane inversion symmetry leads to the splitting of doubly degenerate states, and therefore, there exist more pairs of neighboring energy subbands. The splitting electronic states are expected to induce more excitation channels and absorption spectrum structures. Most of the energy bands present parabolic dispersions, while the $(v_1,c_1)$ and $(v'_1,c'_1)$ energy subbands around the half filling exhibit the oscillating and crossing behaviors. In general, the band-edge states in various energy subbands seriously deviate from the $K$ point. They will be responsible for the main features of van Hove singularities in the density of states and optical absorption spectrum. Most importantly, the strong hybridizations exist between the neighboring 1D energy subbands, as clearly identified from the wave functions in Fig. 4 [discussed later]. As a result, it needs to redefine the energy subband  indices in the ordering of $(v_1,c_1)$, $(v'_1,c'_1 )$, $(v_1,c_2)$, $(v'_2,c'_2)$, ... etc.\\

The spatial distributions of Bloch wave functions belong to the unusual standing waves, especially for the DW regions. The symmetric and antisymmetric standing waves in a unit cell thoroughly disappear under the modulation of stacking configuration, as clearly indicated in Figs. 4(a)-4(h). Apparently, the oscillation modes are totally changed under the stacking modulation, as identified from a detailed comparison between Figs. 4 and 3. The number of nodes is fixed for all the four sublattices. Furthermore, it grows with the increasing state energies, two and four zero points of the [$v_1$, $c_1$], [$v'_1$, $c'_1$], [$v_2$, $c_2$] [$v'_2$, $c'_2$] energy subbands, respectively shown in Figs. 4(a)-4(b), 5.4(c)-5.4(d), 5.4(e)-5.4(f), and 5.4(g)-5.4(h). The subenvelope functions might be vanishing at the specific wave vectors in the AB, DW, BA or DW regions. Usually, the weights within DWs are relatively small, compared with those in the normal stacking regions. When the weight is large for one region, it becomes small for the others. This clearly illustrates the drastic changes for the bilayer graphene with DWs. Moreover, the tight-binding functions on four sublattices roughly have a linear superposition relationship in the AB and BA stacking regions, being sensitive to the modulated DW regions. For conduction band at the K point, the subenvelope functions on the A$^1$ and A$^2$ [B$^1$ and B$^2$] sublattices are roughly in-phase for the normal stacking region with the dominant weight and out-of-phase for another one. These simple relations might be seriously destroyed in the domain walls. Furthermore, those on the [A$^1$, B$^1$] and [A$^2$, B$^2$] sublattices present the approximately symmetric and anti-symmetric distributions, respectively. On the other hand, the K states of valence energy subbands exhibit the opposite behaviors. The above-mentioned irregular standing waves will strongly affect the optical vertical excitations, e.g., the absence of specific optical selection rules. \\

Electronic energy spectra are greatly diversified by a uniform perpendicular electric field, as clearly illustrated in Fig. 2(c). That is, the non-uniform environment is largely enhanced by $V_z$. A gate voltage across bilayer graphene can create the layer-dependent Coulomb potentials or on-site energies. It leads to the breaking of mirror symmetry about the $z=0$ plane, and so does the inversion symmetry on the $(x,y)$ plane. For a normal bilayer AB stacking, a gate voltage opens a band gap \cite{Oostinga;151}. However, the geometry-modulation system might present the semimetallic behavior as a result of the finite, but the low density of states at the Fermi level [discussed later in Fig. 6]. The rich and unique energy dispersions frequently appear, especially for those near the Fermi level. Most of the energy levels are repelled away from $E_F=0$ by $V_z$, where certain energy dispersions might be rather weak within a finite $k_y$-range. Moreover, the oscillatory valence and conduction subbands, the left- and right-hand pairs of oscillatory subbands, being very close to the Fermi level, linearly intersect near the K point. This crossing phenomenon results in the creation of holes and electrons simultaneously. That is, two pairs of the linearly crossing subbands occurs near $E_F$, thus leading to a special van Hove singularity in density of states. The semi-metallic property is in great contrast with the semiconducting behavior in the gated bilayer AB stacking. \\

As for valence and conduction wave functions, the electric potential and geometric asymmetry make them become highly complicated, as clearly displayed in Figs. 5(a)-5(h). The $V_z$-enhanced abnormal standing waves exhibit the irregular features in the oscillatory forms, amplitudes, zero-point numbers, and relations among four sublattices. In general, there are no analytic sine/cosine waves suitable for such random spatial distributions. The amplitudes could appear at any AB/DW/BA/DW regions in the absence of a concise rule. The number of zero points does not grow with state energies monotonously; furthermore, it might be identical or different on the top and bottom layers. For example, the $v_1$ [$v'_1$] energy subband at the K point, as shown in Fig. 5(a) [Fig. 5(c)], presents the 4- and 4-zero-point [4- and 6-zero-point] subenvelope functions on the first and second layers, respectively. Of course, a simple linear combination of the $[A^1, A^2]$- and $[B^1, B^2]$-dependent tight-binding functions is thoroughly absent, while it almost remains similar for the $[A^1, B^1]$ and $[A^2, B^2]$ sublattices. Apparently, the above-mentioned features of Bloch wave functions are expected to induce very complex optical excitation channels. \\

Specifically, the subenvelope functions of the Fermi-momentum states, which are situated in the linearly intersecting valence and conduction subbands [Fig. 2(c)], are worthy of a closer examination. There exist four Fermi momenta, namely, $k_{F1}$, $k_{F2}$, $k_{F3}$, and $k_{F4}$, in which the former [latter] two consists of a pair due to the valence [conduction] hole [electron] states. Apparently, their wave functions exhibit the unusual phenomenon, as clearly displayed in Figs. 6(a)-6(d). That is to say, they are mostly localized in/near a certain region of Domain wall, e.g., the $k_{F1}$/$k_{F4}$ [$k_{F2}$/$k_{F3}$] state corresponding to the second [first] domain  Moreover, the localization behavior is accompanied with a plateau-structure density of states across $E_F$ [discussed in Fig. 7]. Apparently, this result suggests that the metallic transport properties might come to exist along the $y$-direction of a narrow domain wall. Such a phenomenon is similar to those in metallic/armchair carbon nanotubes\cite{5JPSJ81;064719,5PCCP18;7573}. The high-resolution STS measurements could directly verify the significant characteristics of electronic energy spectra and wavefunctions under the different gate voltages. \\

The significant characteristics of electronic structures are directly revealed as various van Hove singularities in the density of states, as clearly displayed in Figs. 7(a)-7(c). A pristine system has a finite, but low magnitude at $E_F$ [Fig. 7(a)], indicating the semi-metallic property. A very weak band overlap leads to a pair of very close shoulder structures across the Fermi level, being consistent with the first-principles calculations \cite{Aoki;123}. Also another two well separated shoulder structures, which comes from the valence and conduction bands of the second pair, appear at the deeper/higher energies, $\sim -0.43$ eV and 0.45 eV. Such structures are due to the extreme points [the local minima and maxima] of 2D energy bands in the energy-wave-vector space. On the other side, the geometry modulation can create a quasi-1D system and thus the dimension-diversified features [Figs. 7(b) and 7(c)]. There are a lot of asymmetric pronounced peaks divergent in the square-root form, mainly owing to 1D parabolic energy subbands. They could be further classified into single- and double-peak structures, in which most of vHSs belong to the latter. The coexistence of the composite structures obviously corresponds to a non-uniform state splitting [Figs. 2(b) and 2(c)]. A finite DOS at $E_F$ in the geometry-modulated system shows the semi-metallic behavior, being accompanied with a pair of rather strong peak structures at $\sim \pm 0.01$ eV. The latter are closely related to the weakly dispersive energy bands near $E_F$. Under a gate voltage, the DOS near $E_F$ becomes a plateau structure. Apparently, it is induced by the linear energy dispersions across $E_F$, as shown in Fig. 2(c). Furthermore, two very prominent peaks appear at $\sim\pm0.1$ eV, mainly due to the weak dispersions generated by the gate voltage. It should be noticed that the $E_F$-dependent densities of states are comparable under the geometric and electric-field modulations. That is, these manipulation methods cannot generate the unusual transitions among the semi-metallic, metallic, and semi-conducting behaviors. However, they have created the important differences in the low-energy van Hove singularities, especially for those across the Fermi level. \\

The geometry- and gate-modulated bilayer graphene systems exhibit the rich and unique absorption spectra. The joint density of states, which is proportional to the number of vertical optical excitations, is very different from one another among three kinds of bilayer systems, as clearly revealed in the inset of Fig. 8. A pristine system is finite at $\omega =0$ (the dashed black curve), reflecting the parabolic band-edge states at $E_F$ [Fig. 2(a)]. It is featureless below $\omega =0.41$ eV, and then has a shoulder structure there due to the 2D first/second valence band and second/first conduction band (details in [Ref\cite{5SciRep9;859}]). JDOS at $\omega =0$ remains almost the same in the asymmetric bilayer systems [the dashed red curve], while there exist a plenty of asymmetric peaks in the square-root form. Such peaks are very prominent, when the valence and conduction band-edge states correspond to the same wave vector, such as, $(v_1,c_1,v_2,c_2)$ around $K$ and $(v_2,c_3)$ at $k_y a\approx2.06$. However, some structures are weak but observable because of the non-vertical relation between them. Specifically, the magnitude of JDOS and the number of peak structures are greatly reduced under the effect of gate voltage, as shown in the inset of Fig. 8. Such results mainly arise from the optical excitations of  band-edge/non-band-edge valence states to non-band-edge/band-edge conduction ones [Fig. 2(c)]. \\

In addition to van Hove singularities in joint density of states, the available optical transition channels, being dominated by the characteristics of subenvelope functions, will co-determine 1D absorption peaks. The AB-stacked bilayer graphene does not show any absorption structures at lower frequency [$\omega <0.3$ eV], as clearly illustrated by the dashed black solid curve in Fig. 8. Both the optical gap and the energy gap are zero. The same result is revealed in the geometry- and gate-manipulated bilayer systems. Apparently, the geometric modulation displays the significant absorption peaks [all the solid curves], only coming from the band-edge states of valence and conduction subbands with distinct index numbers. The $v_n$ to $c_n$ [$v'_n$ to $c'_n$] optical excitations, which are due to the same pair of valence and conduction bands, are forbidden. The main mechanism is the linear symmetry/anti-symmetric superposition of the [A$^1$, A$^2$] and [B$^1$, B$^2$] sublattices, as discussed in Fig. 4(a)-4(h). The first absorption peak, corresponding to [$v'_1$, $c_2$] at $k_y a=2.06$, comes to exist at 0.049 eV. Furthermore, the second, third, fourth and fifth significant absorption channels are, respectively, related to $(v_2,c_5)$, $(v_6,c_3)$, $(v_2,c_7)$ and $(v_7,c_3)$. Of course, there exist many weak absorption structures in between them, and certain very strong peaks being closely related to multi-channel optical excitations [Fig. 8]. Apparently, there are no specific optical selection rules, being thoroughly different the edge-dependent ones in 1D graphene nanoribbons \cite{5PCCP18;7573}. The main reason is that the 1D quantum confinement can create the well-behaved standing waves\cite{5JPSJ81;064719,5PCCP18;7573}. [Fig. 8]. \\

The intensity, frequency, number, and form of optical special absorption structures are very sensitive to the changes in the width of DW and the perpendicular electric-field strength. In general, a simple relation between these optical features and the modulation width is absent at lower $\omega$'s [Fig. 8]. However, the red-shift phenomenon in the increase of width appears at the larger ones. Apparently, the first absorption structure might be replaced by another excitation channel during the variation of DW width. On the other side, the spectral absorption functions for a specific geometry-modulated bilayer graphene clearly present a regular variation under various gate voltages [Fig. 9]. That is to say, the reduced intensity and the enhanced number of absorption structures occur in the increment of $V_z$. This directly reflects the gradual changes in energy dispersions, band-edge states and wave functions with the gate voltages. It should be noticed that the band-edges states near the Fermi level might exhibit a sharp change, strongly affecting the threshold optical excitations. Certain channels become apparent during the variation of $V_z$, as indicated in Fig. 9. \\

The periodical boundary condition in the asymmetry-enriched bilayer graphene is responsible for the rich 1D electronic and optical properties, being in sharp contrast with 2D behaviors in sliding\cite{5SciRep4;7509} and twisted \cite{5PRB87;205404} systems. The latter two exhibit two pairs/more pairs of 2D energy bands composed of the 2p$_z$ orbitals, as clearly displayed along the high-symmetry points of the first Brillouin zone. The wave-vector-dependent energy dispersions belong to the linear, parabolic oscillatory and partially flat forms, in which the first ones are the well-known vertical/non-vertical Dirac cones in the AA/AA$^\prime$ stacking. Their band-edge states, which are the critical points in the energy-wave-vector space, respectively, correspond to the Dirac points, the extreme and saddle points, the highly-degenerate states and the effective 1D constant-energy loops. They present the following vHSs in density of states: the V-shape structures, the shoulder structures $\&$ logarithmically symmetric peaks, delta-function-like peaks, and square-root asymmetric peaks from 1D band edges. These five kinds of special structures are further revealed in the optical absorption spectra. An electric field could create optical gaps/energy gaps in most of bilayer graphene systems.\\

The experimental measurements could verify the predicted band structures, densities of states, and optical absorption spectra. The high-resolution angle-resolved photoemission spectroscopy [ARPES; details in \cite{5NatMat12;887,5ACSNano6;6930,5NanoLett17;1564,5PRB78;201408,5JAP110;013725,5PRB73;045124}] is the only experimental instrument able to directly examine the wave-vector-dependent occupied electronic states. The measured results have confirmed the feature-rich band structures of carbon-related sp$^2$-bonding systems. Graphene nanoribbons are identified to possess 1D parabolic energy subbands centered at the high-symmetry point, accompanied by an energy gap and non-uniform energy spacings \cite{5ACSNano6;6930}. Recently, a lot of ARPES measurements are conducted on few-layer graphenes, covering the proof on the linear Dirac cone in the monolayer system\cite{5NanoLett17;1564,5PRB78;201408,5JAP110;013725}, two low-lying parabolic valence bands in bilayer AB stacking\cite{5NanoLett17;1564}, the coexistent linear and parabolic dispersions in symmetry-destroyed bilayer systems\cite{5NatMat12;887}, one linear and two parabolic bands in tri-layer ABA stacking\cite{5NanoLett17;1564,5PRB78;201408}, the partially flat and sombrero-shaped and linear bands in tri-layer ABC stacking\cite{5NanoLett17;1564,5PRB78;201408}. The Bernal [AB-stacked] graphite possesses the 3D band structure, with the bilayer- and monolayer-like energy dispersions, respectively, at $k_z = 0$ and 1 [K and H points in the 3D first Brillouin zone]\cite{5PRB73;045124}. The ARPES examinations on the geometry- and electric-field-modulated bilayer graphene systems could provide the unusual band structures, such as, the splitting of electronic states, parabolic/oscillatory/linear energy dispersions near the Fermi level, wave-vector-dependent, band-edge states, and semi-metallic properties. These directly reflect the composite effects due to the irregular stacking/the complicated hopping integrals and Coulomb potentials. \\

Up to date, four kinds of optical spectroscopies, absorption, transmission, reflection, and Raman scattering spectroscopies, are frequently utilized to accurately explore vertical optical excitations. Concerning the AB-stacked bilayer graphene, their measurements have successfully identified the $\sim$0.3-eV shoulder structure under zero field\cite{5Nature459;820}, the $V_z$-created semimetal-semiconductor transition and two low-frequency asymmetric peaks\cite{5Nature459;820}, the two rather strong $\pi$-electronic absorption peaks at the middle frequency, specific magneto-optical selection rule for the first group of Landau levels\cite{5PRL101;267601}, and linear magnetic-field-strength dependence of the inter-Landau-level excitation energies\cite{5PRL101;267601}. Similar verifications conducted on trilayer ABA stacking cover one shoulder at $\sim$0.5 eV, the gapless behavior unaffected by gate voltage, the $V_z$-induced low-frequency multi-peak structures, several $\pi$-electronic absorption peaks, and monolayer- and bilayer-like inter-Landau-level absorption frequencies\cite{5NatPhys7;944}. Moreover, the identified spectral features in trilayer ABC stacking are two low-frequency characteristic peaks and gap opening under an electric field\cite{5NatPhys7;944}. The above-mentioned optical spectroscopies are worthy of thoroughly examining the vanishing optical gaps in stacking-modulated bilayer graphenes, prominent asymmetric absorption peaks, absence of selection rule, forbidden optical excitations associated with the linear relations in the $[A^1, A^2]$ $\&$ $[B^1, B^2]$ sublattices, and drastic changes/the regular variations in absorption structures due to the modulation of DW width/gate voltage. \\

\subsection{Significant Landau subbands}\label{ch5.2}

The non-uniform magnetic quantization in stacking-modulated bilayer graphene systems is fully explored by the generalized tight-binding model. The various interlayer atomic interactions and the magnetic field are included in the calculation without the low-energy perturbation. The quasi-1D Landau subbands, which are created by the periodical AB/domain wall/BA structure, exhibit the partially flat and oscillatory dispersions with or without the anti-crossing phenomena. The greatly reduced state degeneracy and extra band-edge states lead to more van Hove singularities in the wider energy ranges. The well-behaved, perturbed and seriously distorted wave functions appear at the specific regions associated with the stacking configurations. The close relations among the geometric symmetries, the band-edge states, and the spatial probability distributions are identified from the delicate analyses. \\

The generalized tight-binding model is delicately developed for a full understanding of the non-uniform magnetic quantization due to the stacking-modulated configuration. Most importantly, there exists a very strong competition between the geometric modulation and magnetic field. This method is reliable for the complicated energy bands with the oscillatory dispersions and the multi-constant loops. The non-homogeneous interlayer and interlayer hopping integrals and the magnetic-field Peierls phases effects are simultaneously included in the diagonalization of the giant magnetic Hamiltonian matrix. The ratio of the periods, which is related to the Peierls phase and the stacking modulation, is characterized by the number of supercell [$N_{sc}$]. For example, concerning (60,10,60,10) stacking-modulated bilayer graphene with a period of 140, the Hamiltonian is a $3360\times 3360$ Hermitian matrix with $N_{sc}=6$ under $B_z=95$ T, in which the number of independent magnetic matrix elements is about 560. The magneto-electronic properties, energy dispersions, density of states, and magnetic wave functions, are investigated in detail. Their close relations are directly linked together by the detailed analysis. A lot of oscillatory Landau subbands are predicted to initiate from the Fermi level, in great contrast with the dispersionless Landau levels in the well-stacked graphene systems\cite{5IOP;CY,5CRCPress;CY}. The theoretical predictions on the significant characteristics of quasi-1D Landau subbands could be verified by the high-resolution STS measurements. The typical methods in generating many Landau subbands and the important differences among them are also discussed. \\

Apparently, the magneto-electronic energy spectra of bilayer graphene systems display the rich features, strongly depending on the stacking configurations. A pristine bilayer AB stacking has two groups of valence and conduction Landau levels under the magnetic quantization of the $\pi$-bonding electronic states. The first group is initiated from the Fermi level, as indicated by the red points in Fig. 10. The well-behaved levels, which are similar to electronic states of an oscillator, are highly degenerate in  the reduced first Brillouin zone of the 2D $(k_x, k_y)$-space. Their quantum numbers are characterized by the zero points of the oscillatory distributions in the dominating sublattice; furthermore, they have a normal ordering according to $n^c=1$/$n^v=0$, 2, 3 ..etc. A very small energy spacing between the $n^v=0$ and $n^c=1$ Landau levels is band gap of $\sim 20$ meV across $E_F=0$ under $B_z=95$ T. The dispersionless Landau levels dramatically changes into the quasi-1D Landau subbands with the partially flat and oscillatory $k_y$-dispersions. The band-edge states, directly determining the significant density of states, correspond to the dispersionless $k_y$ states, and the locally extreme states [the local minima and maxima]. When the magneto-electronic states present the almost dispersionless behavior within a certain $k_y$-range, they could be regarded as the quasi-Landau-level states. This is also identified from the well-behaved magnetic wave functions [discussed later in Fig. 11] . For each $(k_x, k_y)$ state, the four-fold degenerate Landau-level states are replaced by the doubly degenerate Landau-subband ones, because the $x$-direction associated with stacking modulation has been specified. Apparently, the asymmetry of electron and hole states is greatly enhanced by the stacking modulation, being attributed to the non-uniform intralayer and  interlayer hopping integrals. The oscillation width grows quickly during the increment of $|E^{c,v}|$, in which the anti-crossing phenomena appear  and thus create more band-edge states. At higher/deeper state energies, the geometry modulation shows the amplified effects, mainly owing to the larger radii of the magnetic cyclotron motions. That is to say, the crossings and anti-crossings of Landau subbands come to exist very frequently. However, this unusual phenomenon is very difficult to observe in a composite external field. \\

Many van Hove singularities of quasi-1D appear as the special structures in density of states, in which they mainly originate from the various band-edge states. The partially flat quasi-Landau-level and the 1D parabolic extreme states, respectively, generate the delta-function-like peaks and the square-root-form asymmetric peaks, as apparently indicated in Fig. 11. A well-behaved AB stacking only presents the regular symmetric peaks\cite{5CRCPress;CY,5IOP;CY}. The height of each peak is related to the Landau-level $(k_x, k_y)$-degeneracy. Two neighboring peaks across the Fermi level, being due to the $n^v=0$ and $n^c=1$ Landau levels of the first group, represent the most important characteristics. The peak positions, the Landau-level energies, clearly display a redshift under the decrease of $n_{sc}$ (the magnetic-field strength). In general, the stacking modulation results in the greatly reduced peak intensities, the significant shifts of the peak positions, and many extra asymmetric peaks in the extended energy ranges. Obviously, these drastic changes directly reflect the main features of quasi-1D Landau subbands, the finite $k_y$-ranges, their energy differences with those of Landau levels, and a plenty of local maxima and minima in the oscillatory band dispersions. The modulation effects are relatively easily observed in the presence of a wider Domain wall/a weaker magnetic field. For example, the density of states under $n_{sc}=6$ [Fig. 11(a)] is more seriously deviated from that of a normal stacking, compared with that of $n_{sc}=12$ [Fig. 11(d)]. \\

The homogeneous and non-homogeneous magnetic quantizations, respectively, presents in the normal and stacking-modulated AB bilayer graphene systems. The former exhibits a lot of fully degenerate Landau level in the reduced first Brillouin zone. Within the chosen gauge, each $k_x, k_y$-state Landau level has four-fold degeneracy except for the freedom degree of spin configuration, in which the magnetic wave functions are localized at 1/6, 2/6, 4/6, and 5/6 of an $B_z$-enlarged unit cell under a zero wave vector. Specifically, the magneto-electronic energy spectrum of bilayer AB stacking, corresponding to the magnetically quantized states of the first pair of valence and conduction bands, is characterized by the normal ordering of $n^v=0$, 2, 3... [$n^c$=1, 2, 3, ...], etc. It is well known that the dominating sublattice depends on the localization center, and its zero point determines the magnetic quantum number. For example, such Landau levels are defined by the dominant B$^1$ sublattice under the 4/6 center, as clearly illustrated in Figs. 12 for $n^c=1$. General speaking, the  quantum-mode differences for A$^1$ $\&$ A$^2$,  B$^1$ $\&$ B$^2$, and A$^1$ $\&$ B$^1$/A$^2$ $\&$ B$^2$ are, respectively, 0, $\pm 2$, and $\pm 1$ [details in Refs\cite{5IOP;CY}]. This directly reflects the intrinsic geometric symmetries, i.e., the equivalence of two sublattices on the same layers, and the different chemical environments of them. \\

The non-uniform intralayer and interlayer hopping integrals have dramatically changed the main features of magnetic wave functions; that is, Landau subbands are quite different from Landau levels in the spatial distributions. The state splitting, as clearly shown in Fig. 10(a), clearly indicates that the neighboring Landau subbands present the 4/6 and 2/6 localization centers, in which the former is chosen for a model study. For example, the subenvelope functions of the $n^c=1$ Landau subband are very sensitive to the change of $k_y$ [the various curves in Figs. 12(a)-12(h)]. The $k_y=0.1$ Landau-subband state [the black curves], even corresponding to the partially flat energy dispersions, exhibits the obvious changes in the amplitudes and modes of magnetic subenvelope functions. The dominating B$^1$ sublattice only reduces its amplitude [Fig. 12(f)], while it remains the one-zero-point antisymmetric distribution under the 4/6 localization center well separated from the domain wall. On the other side, the A$^1$-, A$^2$- and B$^2$-related subenvelope functions, respectively, display the drastic transformations: (I) the great enhancement of amplitude and the zero-point change from 0 to 1, especially for the domain-wall region [Fig. 12(e)], (II) the variation of the well-behaved symmetric distribution without zero points into the oscillatory one [Fig. 12(g)] and (III) the emergence of the one-zero point subenvelope function [Fig. 12(h)]. The above-mentioned unusual behaviors quickly grow as $k_y$ increases from zero to 0.12, since the magnetic localization center approaches, enters  and then crosses the first stacking-modulation region. That is, the enhanced competition between the magnetic field and stacking modulation is responsible for the strong $k_y$-dependence on the main features of the sublattice-dependent subenvelope functions. \\

For $k_y$ from $0$ to $0.05$, the $n^c=1$ Landau subband has a weak parabolic energy dispersion [the black curves in Fig. 10(b)], being different from the normal dispersionless ones. Apparently, the dramatic changes, as clearly indicated in Figs. 12(a)-12(d), cover the obvious deviation of localization center from the domain-wall region, the significant transformation between the B$^1$- and A$^2$-sublattice dominance, the drastic variation of the oscillating modes in these two sublattices, and the reduced contributions in the A$^1$ $\&$ B$^2$ sublattices. According to the subenvelope functions of the whole $k_y$-dependent Landau subband states, there are no simple relations between two any sublattices [discussed earlier in the pristine case], e.g., the absence of the same zero-point number for A$^1$ and A$^2$ sublattices. That is to say, the hexagonal symmetry and the AB stacking symmetry are thoroughly broken by the geometric modulation thoroughly the non-uniform intralayer and interlayer hopping integrals. It should be noticed that the four sublattices would have the comparable amplitudes in the increase of energy/quantum number. In addition, the effective-mass approximation is not unsuitable in solving the unusual magnetic subenvelope functions, since each Landau-subband state might be the superposition of various Landau-level modes. \\

The anti-crossings of two distinct Landau-subband states frequently appear in the magneto-electronic spectra, as shown in Fig. 10(a). The specific anti-crossing, which is due to the $n^{v}=1\sim n^{v}=4$ Landau subbands at smaller wave vectors $k_{y}\sim0.02\to0.04$ and $E^c\sim -0.2\to-0.5$ eV, is chosen for a clear illustration [the red curves in Fig. 10(c)]. When wave vectors are well separated from the anti-crossing center, the former and the latter are approximately characterized by the well-behaved zero points of the dominating B$^1$ sublattice. During the variation of wave vector from $k_y=0.02$ to 0.04, the magnetic subenvelope functions present the gradual alternation, the drastic change corresponding to the highly distorted oscillations across the anti-crossing center, and the dramatic transformation changing into another mode. The fourth and the third Landau-subbands states [the upper and lower branches], respectively, become the three- and four-zero-point oscillation modes after the unusual anti-crossing. It can be deduced that each Landau-subband state consists of the major and minor modes in the non-anti-crossing regions, where the latter might belong to $n^{c,v}\pm 1$, $\pm 2$, ... etc. Apparently, this leads to a unique phenomenon that all the Landau subbands anti-cross one another in the $k_y$-dependent energy spectrum [Fig. 10(a)]. \\

In addition to the stacking modulation in few-layer graphenes, the 1D graphene nanoribbons and 3D graphites also exhibit the quasi-1D Landau subbands in the presence of a uniform perpendicular magnetic field. However, their key features are quite different from one another, e.g., the $B_z$-dependent energy spectra, van Hove singularities, and magnetic wave functions. While the ribbon width is much longer than the magnetic length, many 1D Landau subbands are initiated near the Fermi level, in which each one consists of dispersionless quasi-Landau-level and parabolic energy dispersion [details in Refs\cite{5CRCPress;CY}]. Their densities of states appear in the delta-function-like forms. The almost well-behaved wave functions of quasi-Landau-levels are localized about the ribbon center; furthermore, the standing waves come to exist for the other states. These characteristics directly reflects the strong competition among the finite-size confinement, edge structures and magnetic field. On the other side, the periodic interlayer hopping integrals play critical roles in the AA-, AB- and ABC-stacked  graphites, respectively, displaying one pair \cite{5CRCPress;CY}, two pairs \cite{5CRCPress;CY} and one pair \cite{5CRCPress;CY} of 3D valence and conduction bands. The 3D magnetically quantized states form many Landau subbands which are composed of 2D dispersionless Landau levels and 1D $k_z$-dependent parabolic energy dispersions. The number of Landau-subband groups is, respectively, one , two and one for simple hexagonal, Bernal and rhombohedral graphites. Some Landau subbands in AA-stacked graphite cross the Fermi level, and their bandwidths are $\sim$1 eV, mainly owing to the strongest interlayer atomic interactions in three graphitic systems. Such 1D Landau subbands generate a plenty of square-root asymmetric peaks in density of states, as observed those for Bernal and rhombohedral graphites. The Landau subbands of AB-stacked graphite possess band widths of about 0.2 eV , and only two of them intersect with $E_F$\cite{5PRB83;121201}. The Landau-subband energy spectrum, corresponding to the K and H points ($k_z$ = 0 and 1), is almost asymmetric and symmetric about $E_F$, respectively. The ABC-stacked graphite presents the Landau-subband bandwidth of $<10$ meV, and only one Landau level is located at $E_F$\cite{5JPCM27;125602}. In general, the magnetic wave functions of Landau-subband states are similar to those of the well-behaved Landau levels. \\

The third method in creating the quasi-1D landau subbands is to introduce the specific composite field, the superposition of a uniform perpendicular magnetic field and the spatially modulated magnetic field/electric field\cite{5OptExp22;7473}, The generalized tight-binding is also suitable for fully exploring the non-uniform quantization phenomena, while the external fields are commensurate to each other. For example, such composite fields in monolayer graphene are predicted to exhibit the lower-degeneracy Landau subbands with the strong energy dispersions and high anisotropy. The 1D characteristics further lead to a plenty of square-root-form asymmetric peaks in density of states. Apparently,  the magnetic wave functions present the seriously distorted spatial distribution and even a dramatic transformation of oscillation modes, depending on the strength and period of a spatially modulated field. This field has very strong effects on the main features of Landau subbands, covering the enhancement in dimensionality, reduction of state degeneracy, variation of energy dispersions, generation of band-edge states, and the changes in the center, width, phase and symmetric/anti-symmetric distribution of the localized quantum modes. However, the anti-crossing phenomena are thoroughly absent. In addition, the above-mentioned quasi-1D magneto-electronic states have been expected to display the rich and unique absorption spectra, e.g., the drastic changes of magneto-optical selection rules\cite{5OptExp22;7473}. \\

The anti-crossing phenomena of magneto-electronic states come to exist in the wave-vector-\cite{5OptExp22;7473}, magnetic-field-\cite{5PRB83;195405}, and electric-field-dependent energy spectra\cite{5CPC184;1821}, where the first, second and third types frequently appear in Landau subbands, Landau levels and both. They mainly come from the similar physical pictures/the identical Van-Born theorem\cite{5PhysZ30;467}, being assisted by the critical mechanisms: the special stacking configurations [e.g., the non-AA configurations; Refs\cite{5CRCPress;CY}], the multi-orbital hybridizations\cite{5IOP;SC}, the non-uniform chemical environments\cite{5IOP;SC}, and the significant spin-orbital couplings\cite{5IOP;SC}. However, the specific relation of quantum numbers [the quantum-number difference of $\Delta n$] between the main and side modes might be different in layered condensed-matter systems with various stacking configurations. For example, only the AB-stacked graphite\cite{5PRB83;121201}, but not the AA-\cite{5CRCPress;CY} and ABC-stacked\cite{5JPCM27;125602} systems, present the frequently Landau-subband anti-crossings between the first and second groups during the variation of $k_z$, with $\Delta n=3$ [details in Refs\cite{5CRCPress;CY}]. Apparently, the $k_z$-decomposed wave vectors play important roles on the interlayer atomic interactions and thus the unusual behaviors, e.g., the bilayer- and mono-layer-like hopping integrals at $k_z=0$ and $\pi$, respectively\cite{5PRB83;121201,5CRCPress;CY}. The similar relations are also revealed in the well-stacked graphene systems for the $B_z$-induced energy spectra, such as, the AB \cite{5PRB83;121201}, ABC\cite{5JPCM27;125602} and AAB stackings\cite{5Carbon94;619}. In addition, the undefined Landau levels, corresponding to non-well-behaved sliding bilayer systems\cite{5SciRep4;7509}, exhibit the continuous intergroup anti-crossings. Specifically, the various gate voltages could be utilized to manipulate the frequent anti-crossings of Landau levels as a result of the splitting Landau levels. Such electric fields, which are, respectively, cooperated with the specific interlayer hopping integrals [layered graphenes; Refs\cite{5OptExp22;7473,5PRB83;195405,5CPC184;1821}], the important spin-orbital interactions [germanene and tinene;\cite{5IOP;SC}, and the complicated intralayer and interlayer hopping integrals [bilayer phosphorene;\cite{5IOP;SC}], present quantum number differences: $\Delta n=3I$, $\pm 1$ and $I$. \\

The measurements of scanning tunneling microscopy are very powerful in exploring the van Hove singularities due to the band-edge states and the metallic/semiconducting/semi-metallic behaviors, They have successfully identified diverse electronic properties in graphene nanoribbons\cite{5PCCP18;7573}, carbon nanotubes\cite{5CRCPress;CY}, few-layer graphene systems\cite{5IOP;SC,5CRCPress;CY}, and graphite\cite{5CRCPress;CY}, The focuses of the STS examinations on the geometry-modulated and gated bilayer graphenes should cover the square-root asymmetric peaks, the major double-peak structures and the minor single-peak ones, and the significant density of states at the Fermi level. Furthermore, a pair of very close shoulder structures just across $E_F$, a finite DOS at $E_F$ accompanied by the prominent valence and conduction asymmetric peaks, and a sufficiently long plateau crossing $E_F$, can distinguish the distinct effects of AB configuration, stacking modulations. and non-uniform hopping integrals $\&$ Coulomb potentials, respectively. Moreover, such STS measurements are very reliable for identifying the uniform and non-uniform magnetic quantization in electronic energy spectra of layered graphene systems. As to the former, the measured tunneling differential conductance directly reflects the structure, energy, number and degeneracy of the Landau-level delta-function-like peaks. Part of the theoretical predictions on the Landau-level energy spectra have been verified by the experimental measurements, e.g., the $\sqrt{B_z}$-dependent Landau-level energy of monolayer graphene\cite{5IOP;SC,5CRCPress;CY}, the linear $B_z$-dependence in AB-stacked bilayer graphene\cite{5IOP;SC,5CRCPress;CY}, the coexistent square-root and linear $B_z$-dependences in trilayer ABA stacking\cite{5IOP;SC,5CRCPress;CY}, and the 2D and 3D characteristics of the Landau subbands in Bernal graphite\cite{5CRCPress;CY,5PRB83;121201}. The experimental examinations on the energy range and number of prominent asymmetric and symmetric peaks of the stacking-modulated bilayer graphene systems can clearly illustrate the significant characteristics of the oscillatory quasi-1D Landau subbands. namely, the partially flat and parabolic dispersions, the oscillation width, and the subband anti-crossing phenomena. \\

The energy-fixed STS measurements can directly map the spatial probability distributions of wave functions in the presence/absence of a uniform magnetic [electric] field. Up to now, they have verified the rich and unique electronic states in graphene-related systems. For example, the well-behaved standing waves, with the specific zero points, could survive in finite-length carbon nanotubes\cite{5Science283;52}. The topological edge states are identified to be created by the AB-BA domain wall of bilayer graphene systems\cite{5NatComm7;11760}. Moreover, the normal Landau-level wave functions, being similar to those of an oscillator, frequently appear in the well-stacked few-layer graphene systems, such as, the $n^{c,v}=0$, and $\pm 1$ Landau levels for the monolayer and AB-stacked bilayer graphenes\cite{5PRL94;226403}. The higher-$n$\cite{5PRB83;121201}, undefined\cite{5SciRep4;7509}, and anti-crossing Landau levels\cite{5PRB90;205434} are worthy of the further experimental examinations. The similar STS measurements are available in examining the current predictions: (I) the drastic changes in the main features of standing waves due to the stacking modulation, (II) the gate-voltage-induced localized states within the Domain walls, directly linking with a plateau-structure density of states across the Fermi level, and (III) the non-uniform magnetic quantization, establishing the direct relations between the band-edge state energies and the distribution range of the regular or irregular magnetic wave functions. \\

\par\noindent {\bf Acknowledgments}

This work was supported in part by the National Science Council of Taiwan,
the Republic of China, under Grant Nos. NSC 98-2112-M-006-013-MY4 and NSC 99-2112-M-165-001-MY3.

\newpage
\renewcommand{\baselinestretch}{0.2}

\begin{figure}
\centering \includegraphics[width=0.9\linewidth]{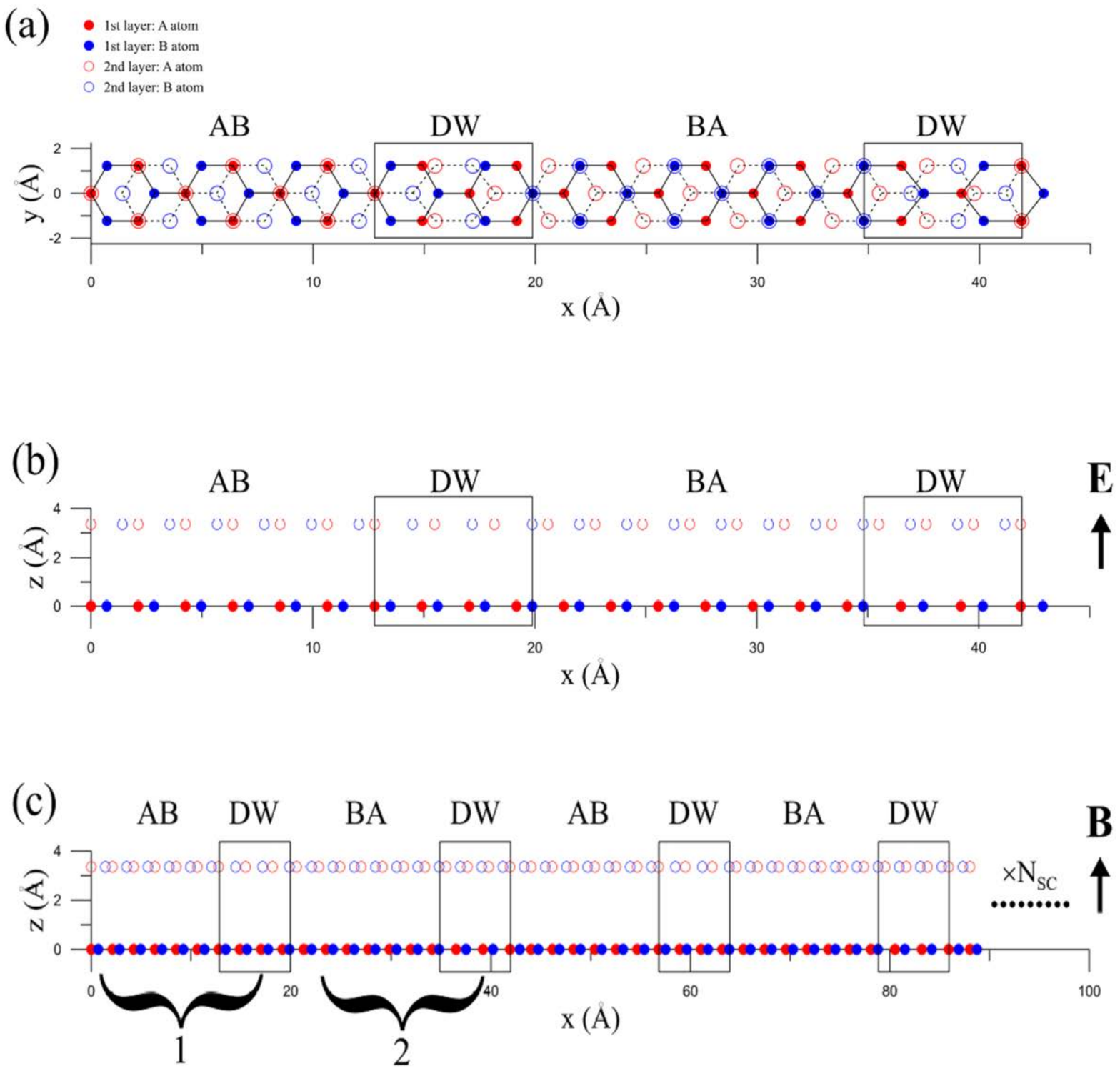}
\begin{center} Figure 5.1: Geometric structures of the stacking-modulated bilayer graphenes associated with the AB configuration under the (a) top, (b) side views with a uniform perpendicular electric field, and (c) an enlarged unit cell in a commensurate magnetic field.
\end{center} 
\end{figure}

\newpage
\begin{figure}
\centering \includegraphics[width=0.9\linewidth]{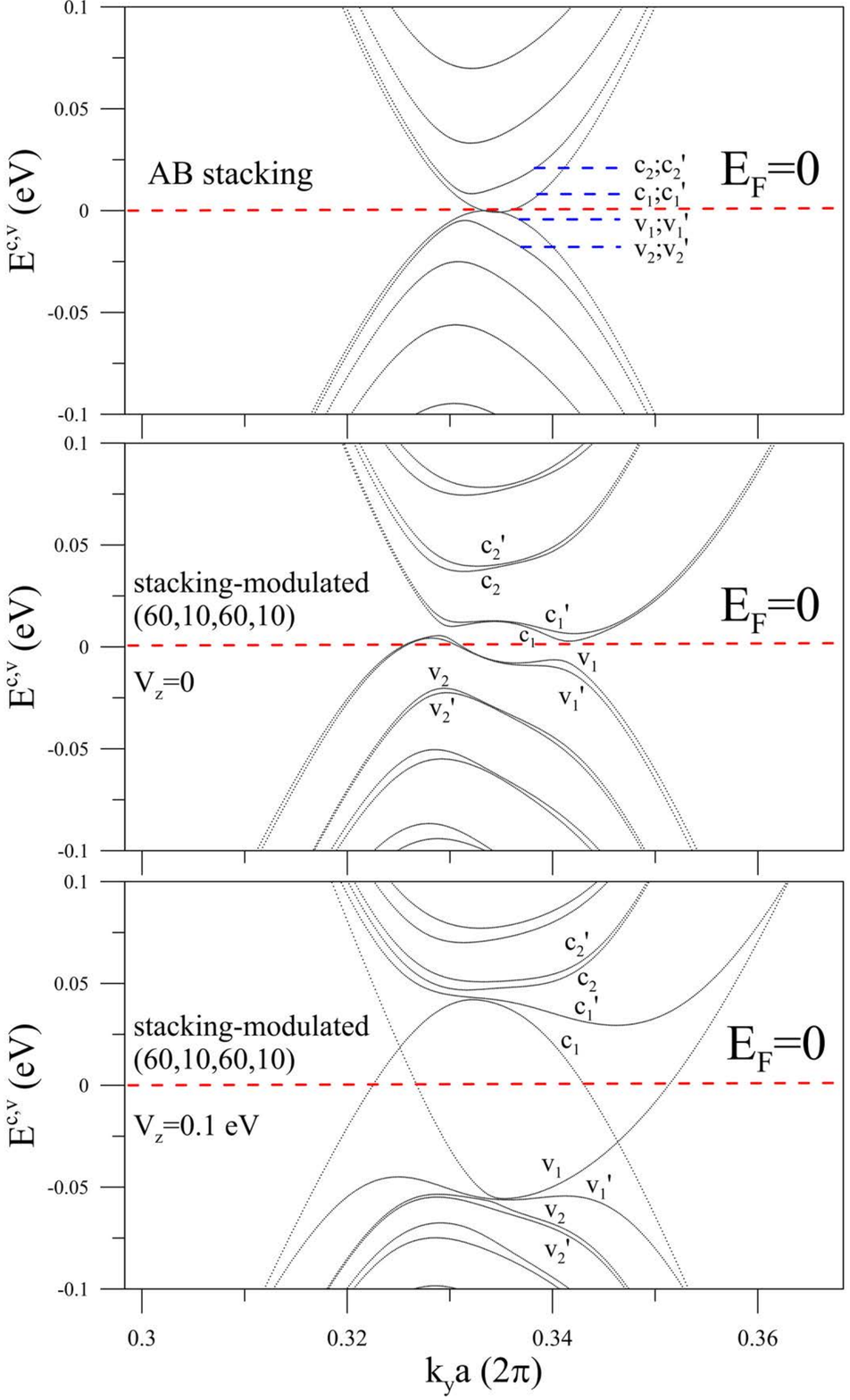}
\begin{center} Figure 5.2: The low-lying energy bands initiated from the $k_ya=2\pi /3$ state: (a) a pristine bilayer AB stacking, (b) the stacking-modulated AB/DW/BA/DW (60,10,60,10), and (c) the geometry- and electric-field-manipulated system under gate voltage of $V_z= 0.1$ eV.
\end{center} \end{figure}
\newpage
\begin{figure}
\centering \includegraphics[width=0.9\linewidth]{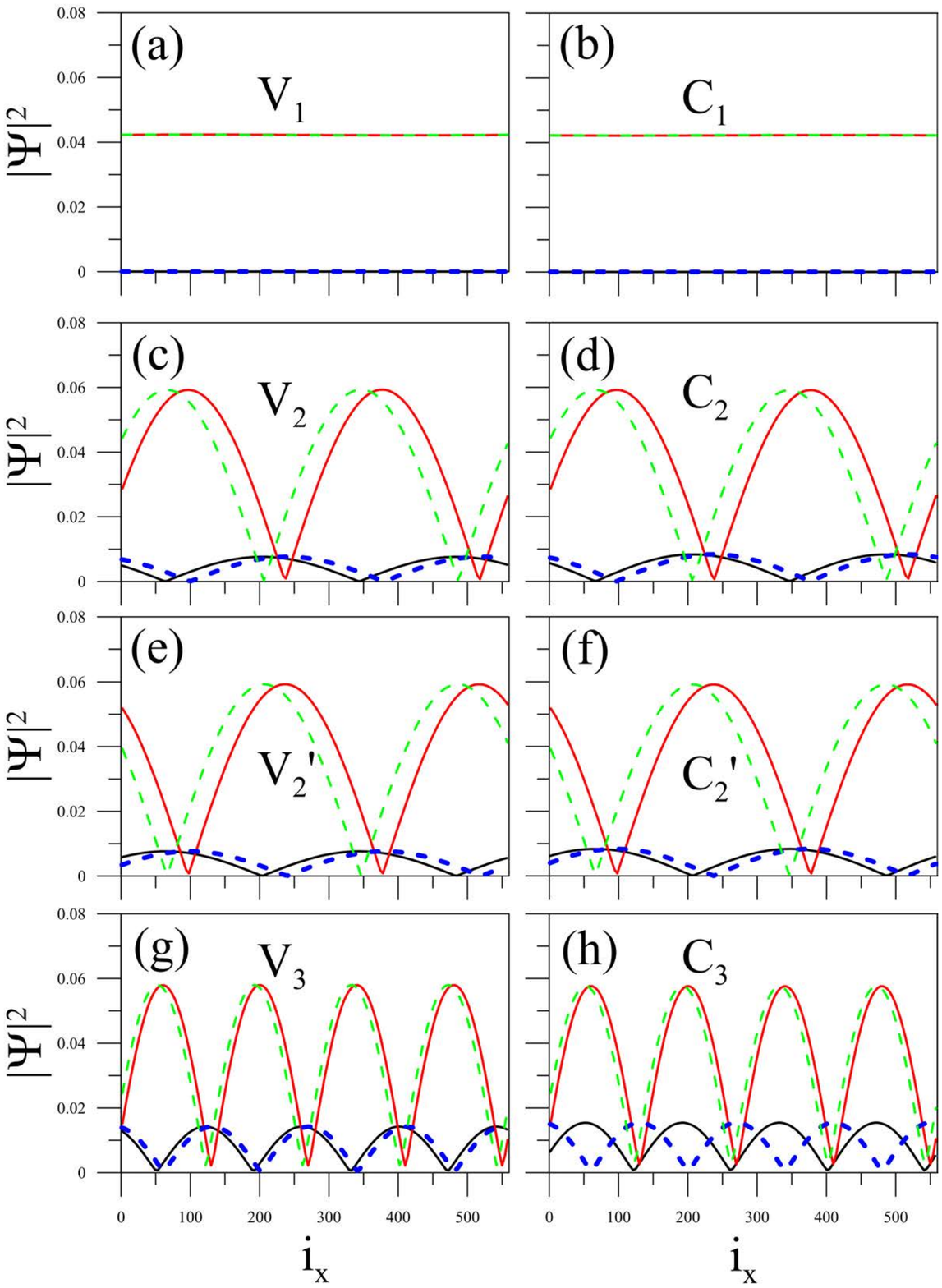}
\begin{center} Figure 5.3: the zero-field subenvelope functions of the AB-stacked bilayer graphene in the $(A^1, B^1, A^2 , B^2)$ sublattices for the K point under an enlarged unit cell: the (a) $v_1$, (b) $c_1$, (c) $v_2$, (d) $c_2$, (e) $v'_2$, (f) $c'_2$, (g) $v_3$, and (h) $c_3$ energy subbands.
\end{center} \end{figure}
\newpage
\begin{figure}
\centering \includegraphics[width=0.9\linewidth]{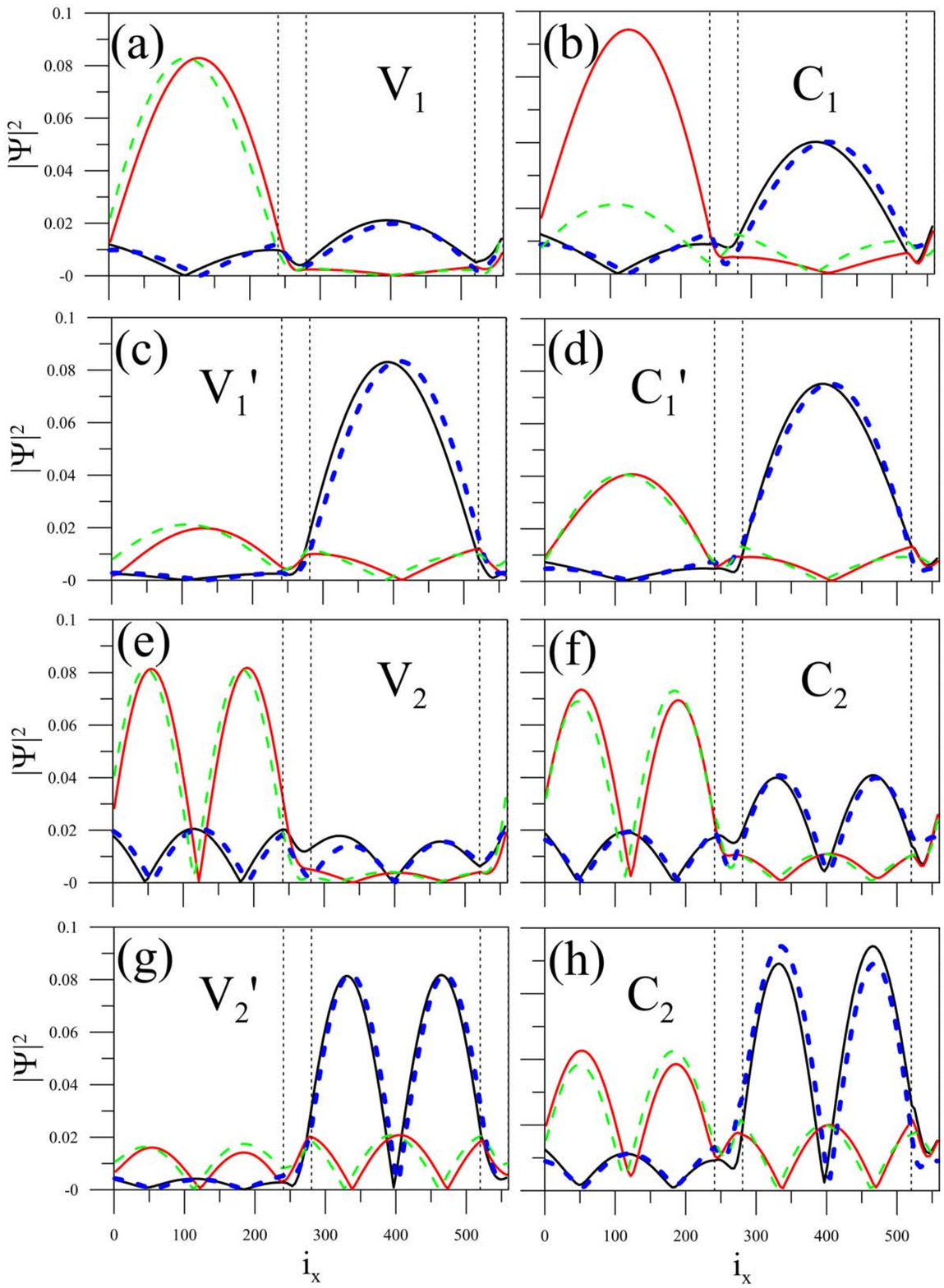}
\begin{center} Figure 5.4: Similar plots as Fig. 5.3, but shown for the AB/DW/BA/DW bilayer graphene at the specific K point of the energy subbands: (a) $v_1$, (b) $c_1$, (c) $v'_1$, (d) $c'_1$, (e) $v_2$, (f) $c_2$, (g) $v'_2$ and (h)) $c'_2$, according to the energy ordering.
\end{center} \end{figure}
\newpage
\begin{figure}
\centering \includegraphics[width=0.9\linewidth]{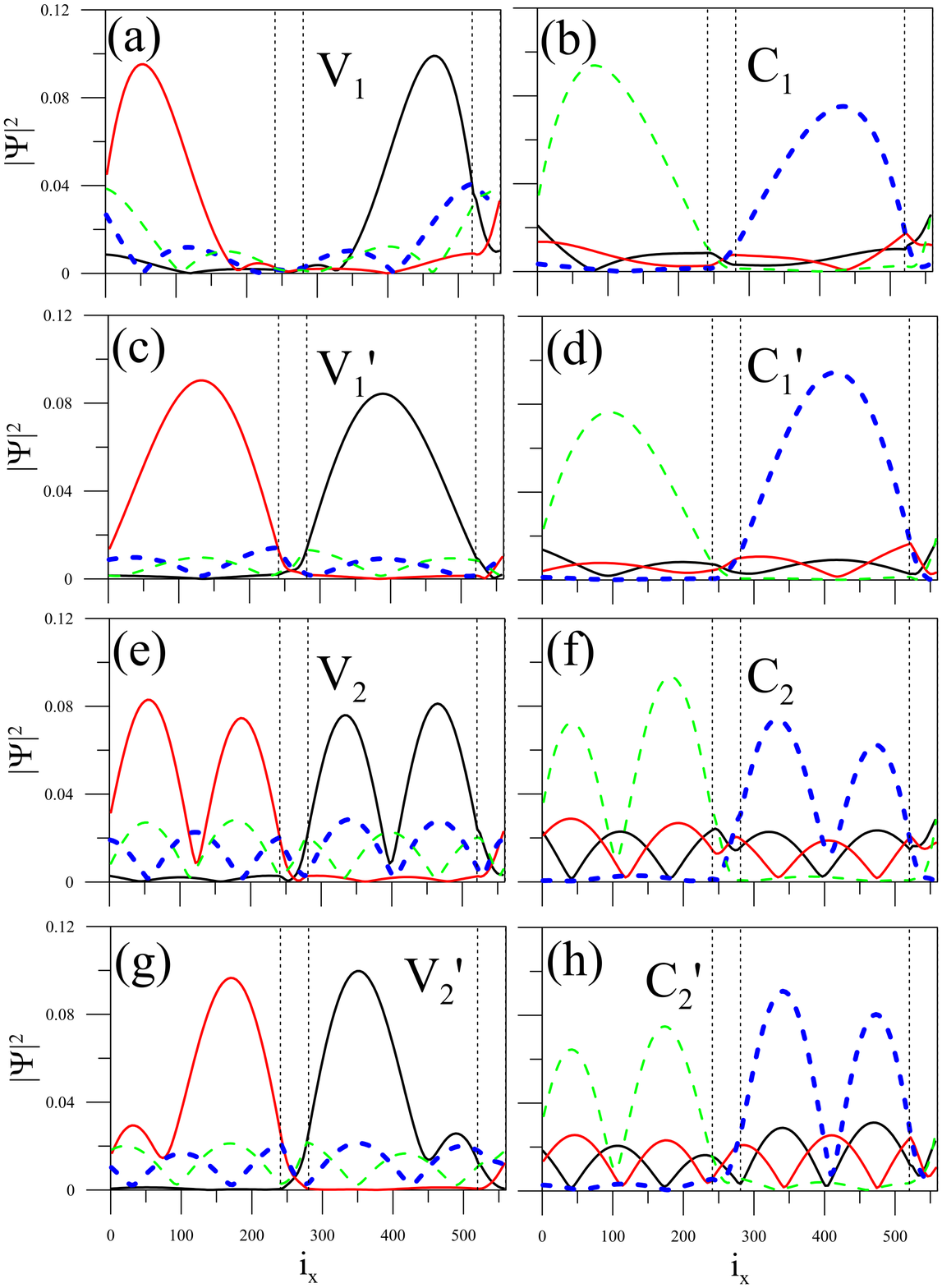}
\begin{center} Figure 5.5: The stacking- and voltage-modulated subenvelope functions on the four sublattices at $V_z=0.1$ eV for the K point in the energy subbands: (a) $v_1$, (b) $c_1$, (c) $v'_1$, (d) $c'_1$, (e) $v_2$, (f) $c_2$, (g) $v'_2$, and (h) $c'_2$.
\end{center} \end{figure}
\newpage
\begin{figure}
\centering \includegraphics[width=0.9\linewidth]{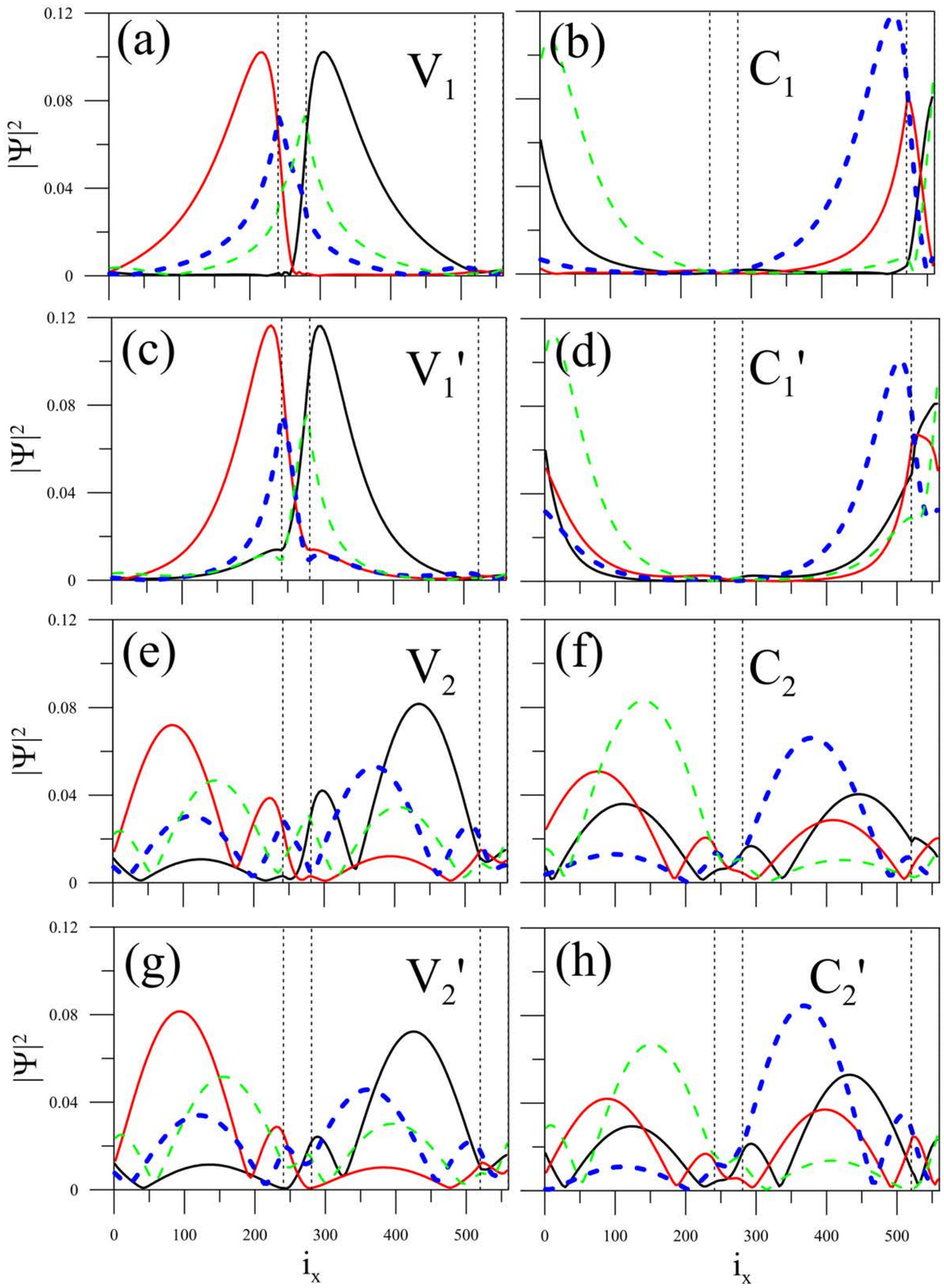}
\begin{center} Figure 5.6: Similar plot as Fig. 5.5, but displayed under for the four momentum states in the linearly intersecting energy subbands.
\end{center} \end{figure}
\newpage
\begin{figure}
\centering \includegraphics[width=0.9\linewidth]{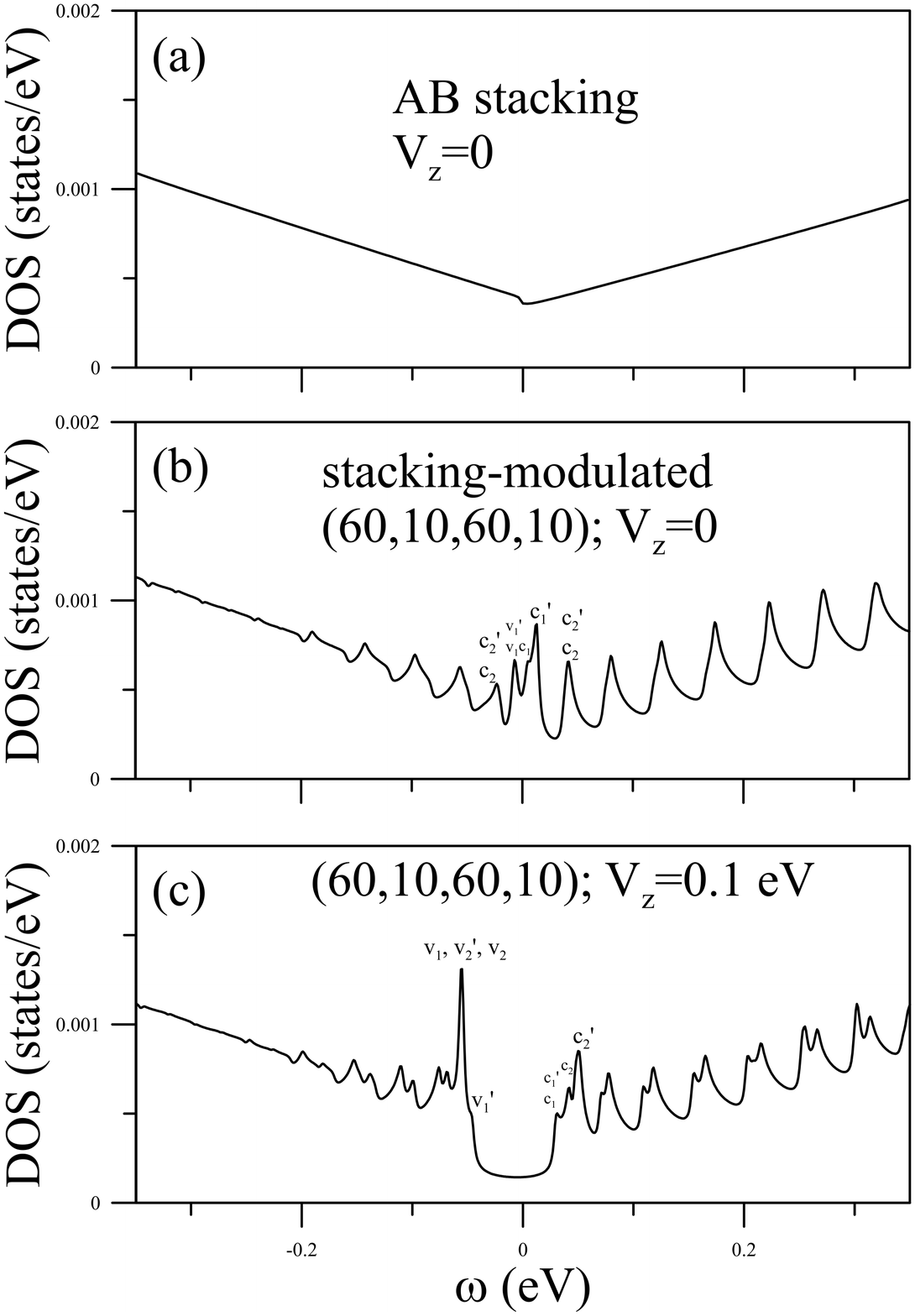}
\begin{center} Figure 5.7: The significant density of states for bilayer graphene systems under (a) a normal AB stacking, (b) a periodical structure of  AB/DW/BA/DW, and (c) a geometric modulation at $V_z=0.1$ eV.
\end{center} \end{figure}
\newpage
\begin{figure}
\centering \includegraphics[width=0.9\linewidth]{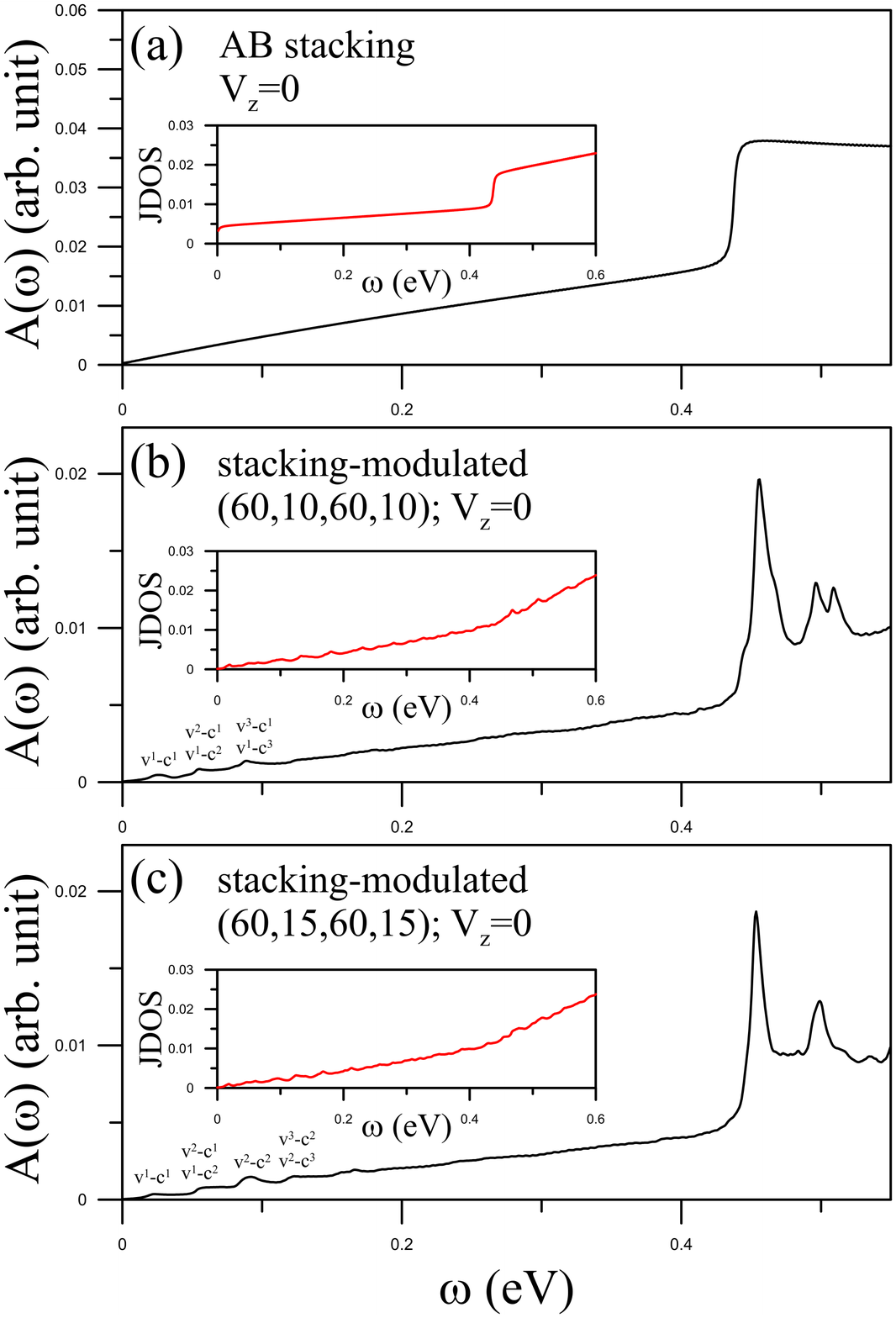}
\begin{center} Figure 5.8: The optical absorption spectra for pristine and stacking-modulated bilayer graphene systems under various domain-wall widths. Also shown in the the inset are the typical joint densities of states.
\end{center} \end{figure}
\newpage
\begin{figure}
\centering \includegraphics[width=0.9\linewidth]{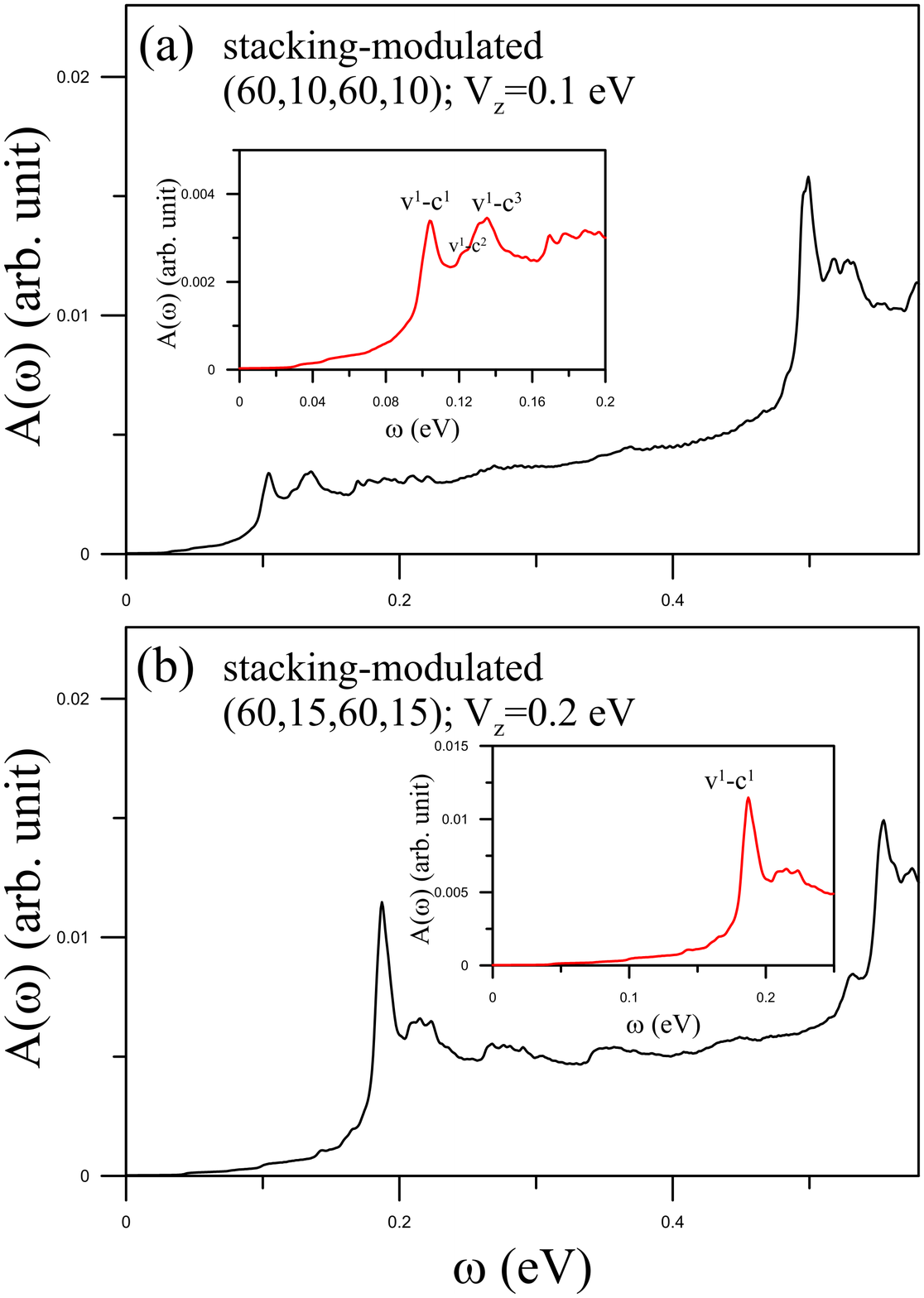}
\begin{center} Figure 5.9: Similar plot as Fig. 5.8, but only shown for stacking modulated systems in the presence of distinct gate voltages. The inset clearly illustrates the detailed absorption structures for typical cases.
\end{center} \end{figure}
\newpage
\begin{figure}
\centering \includegraphics[width=0.9\linewidth]{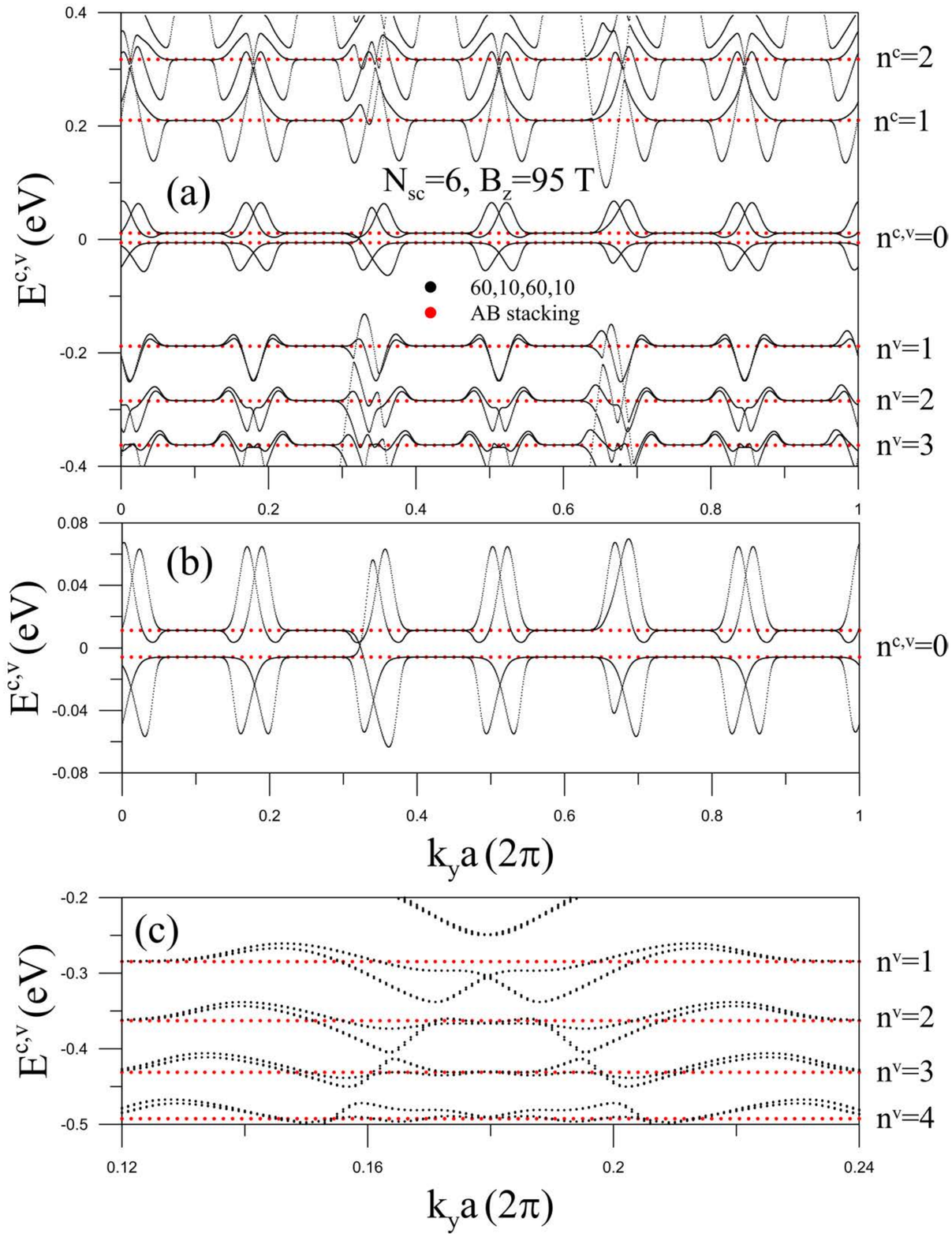}
\begin{center} Figure 5.10: (a) The low-lying Landau subbands of the stacking-modulated bilayer graphene under $B_z=95$ T and $N_{sc}=6$, and the enlarged energy dispersions (b) near the Fermi level $\&$ (c) for certain anti-crossing phenomena.
\end{center} \end{figure}
\newpage
\begin{figure}
\centering \includegraphics[width=0.9\linewidth]{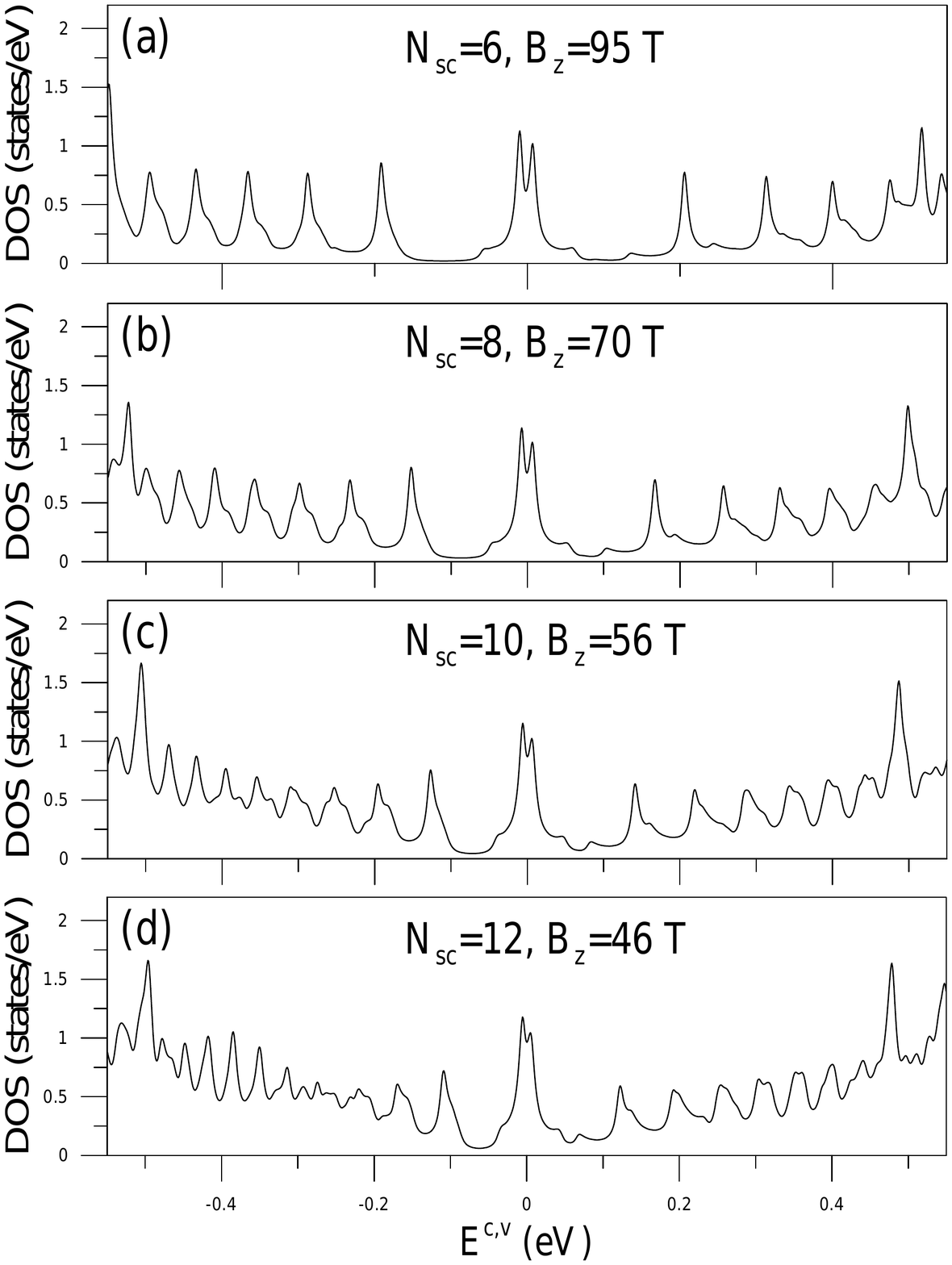}
\begin{center} Figure 5.11: The low-energy density of states for the various domain-wall widths: (a) $N_{sc}= 6$, (b) 8, (c) 10 and (d) 12.
\end{center} \end{figure}
\newpage
\begin{figure}
\centering \includegraphics[width=0.9\linewidth]{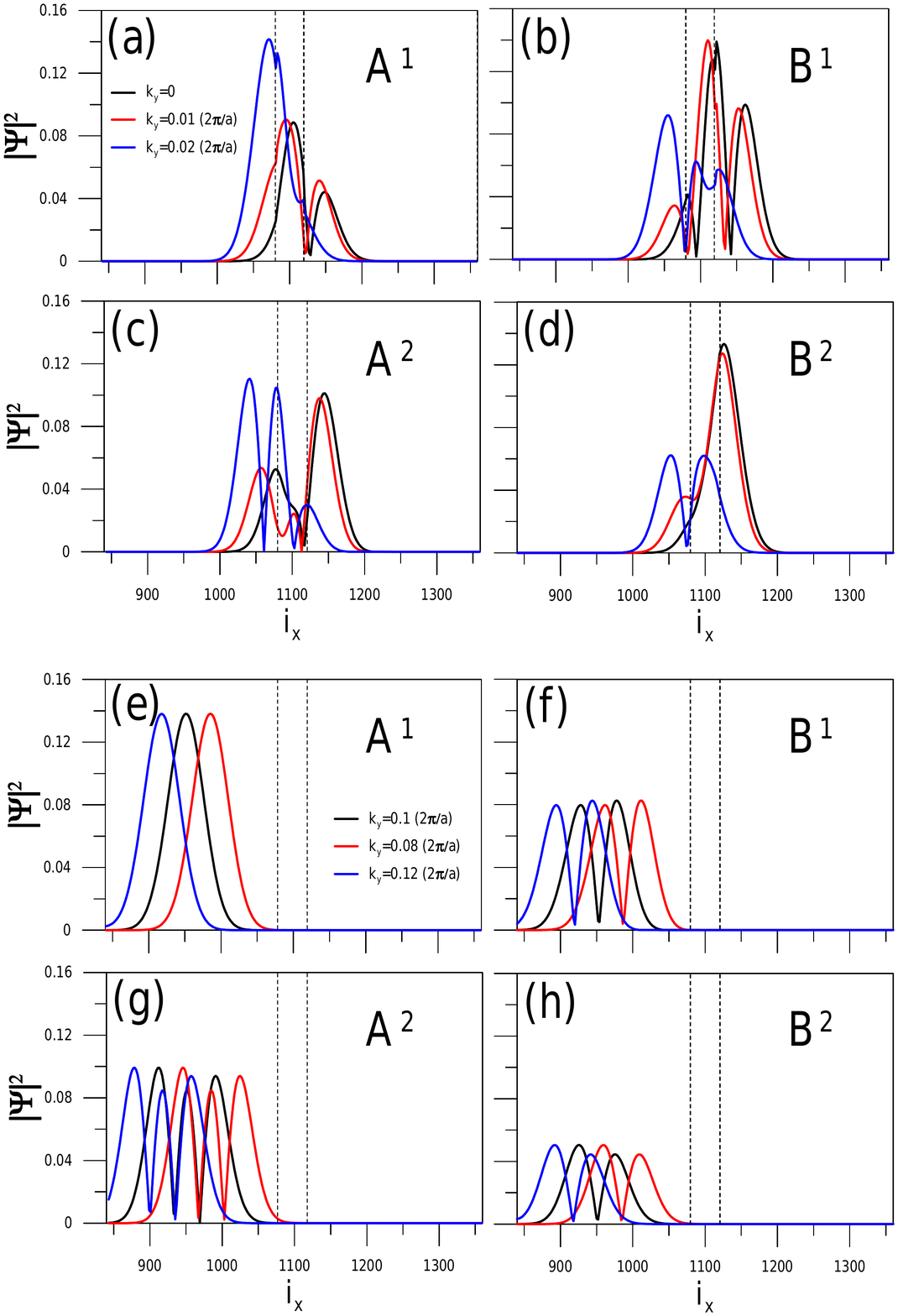}
\begin{center} Figure 5.12: The subenvelope functions for the $n^c=1$ Landau subband, with the localization center near 4/6, on four  sublattices at smaller wave vectors: (a) B$^1$, (b) A$^1$, (C) A$^2$ $\&$ (d) B$^2$, and under larger ones: (e) B$^1$, (f) A$^1$, (g) A$^2$ $\&$ (h).B$^2$.
\end{center} \end{figure}
\newpage
\begin{figure}
\centering \includegraphics[width=0.9\linewidth]{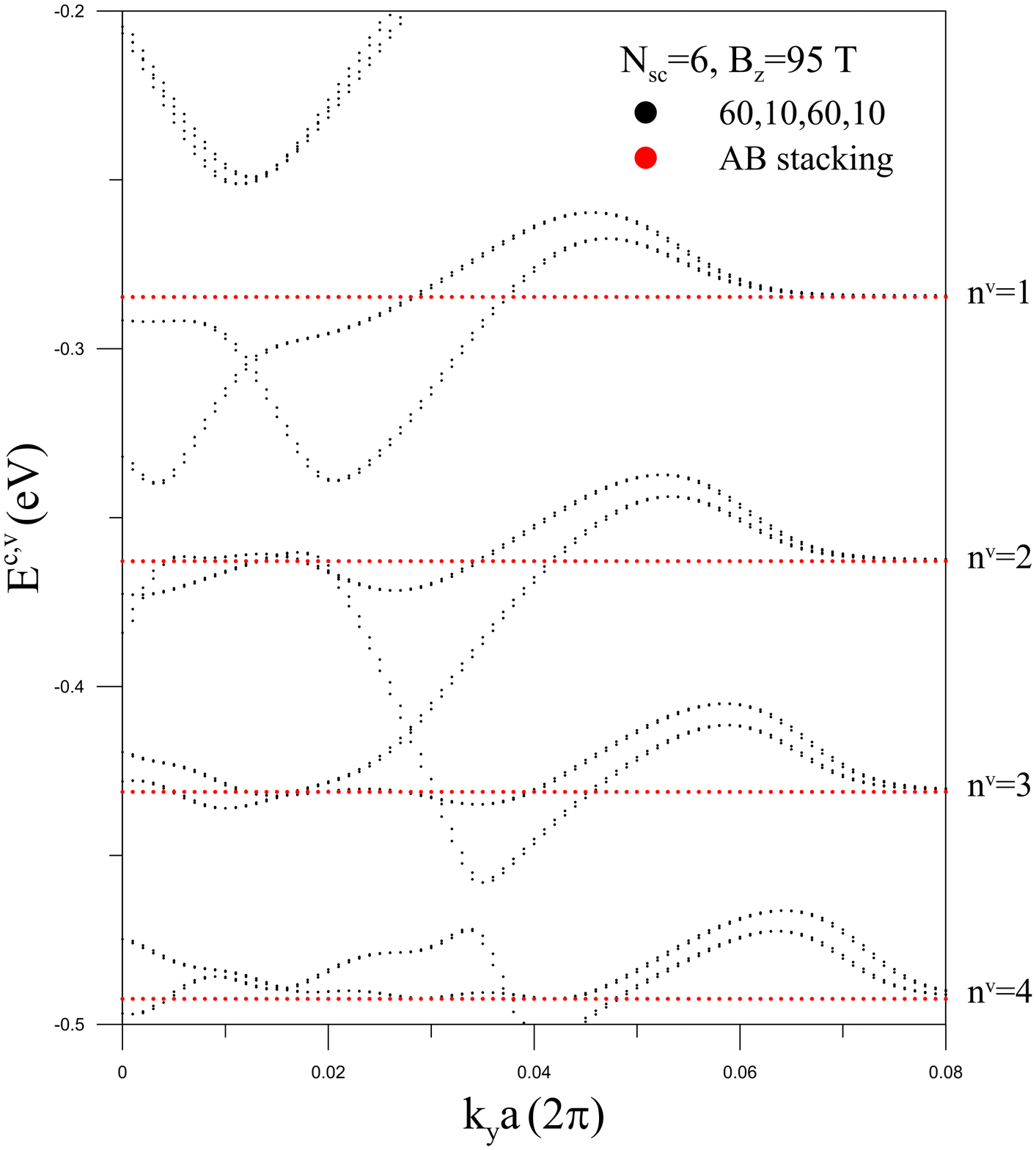}
\begin{center} Figure 5.13: The specific anti-crossing due to the $n^{v}=1\sim n^{v}=4$ Landau subbands at smaller wave vectors and $E^v\sim -0.2\sim-0.5$ eV for the (a) upper and (b) lower branches.
\end{center} \end{figure}

\end{document}